\documentclass[twocolumn,showpacs,preprintnumbers,amsmath,amssymb,pra]{revtex4}

\usepackage{graphicx}

\begin{document}

\title{Single Atom Detection With Optical Cavities}

\author{R.Poldy, B.C.Buchler, and J.D.Close}
\email{john.close@anu.edu.au}
\affiliation{Australian Centre for Quantum Atom Optics, Australian National University, ACT 0200, Australia}

\date{\today}

\begin{abstract}
We present a thorough analysis of single atom detection using optical cavities.  The large set of parameters that influence the signal-to-noise ratio for cavity detection is considered, with an emphasis on detunings, probe power, cavity finesse and photon detection schemes.  Real device operating restrictions for single photon counting modules and standard photodiodes are included in our discussion, with heterodyne detection emerging as the clearly favourable technique, particularly for detuned detection at high power.
\end{abstract}


\maketitle

\section{\label{sec:intro}Introduction:\protect\\}
The experimental realization of Bose-Einstein condensation \cite{Ensher:1995p3052,vanDruten:1995p3004,Bradley:1995p1825} and the coherent outcoupling of atoms from a condensate to form an atom-laser beam \cite{Mewes:1997p22,LeCoq:2001p1913, Hagley:1999p2175, Bloch:1999p12, Robins:2006p2085} have paved the way for the emerging field of quantum atom optics, where atoms play the role analogous to photons in traditional quantum optics experiments \cite{Moelmer:2003p2815}. This new field holds great promise for precision measurement using atoms \cite{RBerman:1997p301} and investigations in fundamental physics \cite{Kasevich:2002p8}. Despite the many similarities between quantum-photon optics and quantum-atom optics, there are fundamental and practical differences. One important practical difference, that is the focus of the present work, is the experimental detection of individual quanta.

Single neutral atoms have been observed and counted using a variety of techniques.  For metastable atoms in highly excited states, such as metastable helium or neon, the internal atomic energy can be used to eject electrons from a metal surface on impact. The electron pulse can be accelerated and detected with good signal-to-noise ratio (SNR) allowing single atom counting \cite{Rasel:2001p2314, Perrin:2007p3873, Yasuda:1996p4812}. Neutral ground-state atoms do not have enough energy for this process. Instead, common detection techniques exploit the interaction of the atom with light. Single atoms have been observed with fluorescence detection \cite{Hu:1994p19,Westphal:1999p5, Alt:2001p3410}, and by measuring the effect on the field in optical \cite{Kimble:1998p6,Munstermann:1999p25} and microwave cavities \cite{Maioli:2005p2562, Brattke:2001p2615}.

\begin{figure}[t!]
\includegraphics[width=\columnwidth]{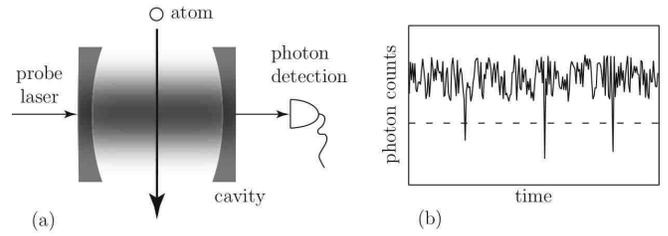}\caption{\label{cavity} Single atom detection with an optical cavity showing (a) schematic diagram of cavity set-up. (b) typical photon counts for a detection event.  The dashed line indicates the threshold for an atom detection event.}
\end{figure}

In recent work, high finesse optical cavities have received much attention as a detection technique \cite{Ritter:2007p2664, Trupke:2007p878}.  Figure~\ref{cavity} is a schematic representation of a possible measurement process.  A shot-noise limited probe laser is transmitted through an empty cavity and the power is measured.  A detection signal is observed when an atom falls through the cavity, interacting with the field and causing, for example, a reduction in transmitted power, as shown in the hypothetical data of Fig.~\ref{cavity}(b).  The interaction may be measured in other ways such as a phase shift in the probe beam, or an increase in transmitted power, as discussed in the following sections.
 
The signal for the cavity detection process has been considered in several studies.  Work by Horak \textit{et al.} investigates optical cavity detection of single atoms using microcavities \cite{Horak:2003p31, Rosenblit:2004p3534,Trupke:2007p878}. The authors use a semiclassical model to analyse the SNR of cavity based detection for a variety of parameters. Pinske \textit{et al.} use a quantum model to calculate the SNR for an atom passing through a cavity  \cite{Pinkse:2000p13}.  They consider a broader range of detection regimes, exploring the effect of the detuning of the probe laser independently from both the empty-cavity resonance (cavity-probe detuning) and bare-atom resonance (atom-probe detuning). Their results however, do not consider the variation of probe power, since they are restricted to the single atom, single photon regime.

There has, to date, been no investigation that thoroughly explores the detection `parameter space'.  Here we analyse the SNR of cavity single atom detection over this space:  we provide contour plots for the SNR as a function of cavity finesse, probe laser intensity, cavity-probe detuning, and atom-probe detuning.  The goal of this paper is to provide detailed information to groups who wish to design and implement cavity based single atom detection that is appropriate for their requirements of quantum efficiency and detection bandwidth \cite{footnote}.

Cavity quantum-electrodynamics (QED) is generally separated into two parameter regimes, those of strong and weak coupling.  These define the relative strengths of coupling within the system and to external reservoir modes.  Strong coupling refers to system dynamics that are largely determined by the atom-field dipole coupling, $g$, and it is this regime that has attracted much of the recent interest in cavity QED.  Consequently, recent investigations and experimental designs have, for the most part, been restricted to regimes that achieve the required strong coupling conditions.  The focus of the present work is not to investigate properties of the strong coupling, but to analyse the cavity detection process in order to determine the best designs and operating conditions for single atom detection.  We show, however, that many of the same features of strong-coupling experiments are necessary for good SNRs in single atom detection.  It is therefore worth clearly defining this regime.

For strong coupling, the exchange of energy within the system occurs on a time scale much shorter than other processes, so that $g>\mathrm{max}[\Gamma, T^{-1}]$, where $\Gamma$ is the set of decoherence rates for the system, and $T$ is the interaction time.  In the atom-cavity system, $\Gamma=\{\gamma,\kappa\}$ with $\gamma$ the rate of decay of the population of the atomic excited state~\cite{footnote3}, and $\kappa$ that of the cavity field.  To achieve strong coupling, it is necessary to ensure the atom-field interaction time, $T$, is long with respect to other system dynamics~\cite{Munstermann:1999p25}.  The atom-field coupling and the cavity decay rates are also important, and the coupling regime is often characterised in more specific terms with two dimensionless parameters; the critical photon number\[m_{0}\equiv\dfrac{(\gamma/2)^{2}}{2g^{2}}\]and the critical atom number\[N_{0}\equiv\dfrac{\gamma\kappa}{g^{2}}=C^{-1}.\]  These values indicate the number of quanta necessary to significantly influence the system.  Strong coupling is usually defined when both $N_{0}$ and $m_{0}$ are less than one.  

The critical photon number is the number of intra-cavity photons necessary to saturate the atom in a resonant transition.  For our purposes, it is not an important value since we are unconcerned with the number of photons needed for a successful detection process.  The critical atom number refers to the necessary number of atoms required to significantly affect the cavity field.  For single atom detection, it is desirable to work with $N_{0}\lesssim 1$, since this implies the presence of a single atom will have a significant influence on the cavity transmission and be easily measured.  $N_{0}$ is often inverted, and referred to as the co-operativity parameter, $C$ \cite{Kimble:1998p6}.

Single-atom detection need not be performed in the strong coupling regime - where both $N_{0}$ and $m_{0}$ are small - as has been highlighted in work by several authors  \cite{Haase:2006p18,Horak:2003p31}.  The main objective is to minimise $N_{0}$ subject to real world experimental constraints.

Short cavities may be used to increase $g^{2}$, which scales as the inverse of cavity volume.  As a result, short cavities lead to a reduction of $m_{0}$.  Cavity length does not, however, influence $N_{0}$, since $\kappa$ scales as the inverse of cavity length.  Motivated by the strong coupling regime, and at times by the complementary requirements of restricted geometry in chip experiments, recent work in cavity QED has tended towards very short cavities, tens to hundreds of micrometers \cite{Mabuchi:1996p165, Hood:1998p5016, Ottl:2005p227, Haase:2006p18, Munstermann:1999p25, Trupke:2007p878}.  Although limiting cavity length will not improve single atom detection, it is possible to reduce the mode volume in ways that do help.  Reducing the beam diameter in the cavity is one such possibility \cite{Haase:2006p18, Trupke:2007p878}.

As well as reducing cavity length, considerable efforts have been made to produce cavities with ultra-high finesse.  This results in reduced cavity linewidth, $\kappa$, without an accompanying reduction in $g$, so is an ideal way to manipulate $N_{0}$.  Whispering gallery mode cavities have reached a finesse of $\mathcal{F}>10^{7}$ \cite{Savchenkov:2007p4873} and open optical cavities with finesses in excess of $\mathcal{F}=3\times 10^{5}$ have been demonstrated using custom built mirrors \cite{Munstermann:1999p25, Ottl:2005p227}.  Custom design and fabrication can be a costly and arduous task, however reasonably high finesses of around $\mathcal{F}\sim 10^{4}$ are within reach even with commercial mirrors \cite{supermirrors}.  Although this is a trade-off in mirror quality for ease and expense of construction, we will show that it is still possible to achieve a good SNR for single atom detection using such a finesse, provided the system is operated in appropriate regions of parameter space.  Determining where these regions are is the motivation for this work, as covered in the following sections.

The layout of the paper is as follows.  Section \ref{model} introduces the atom-cavity field model for an ideal detector using direct photon counting. The SNR in the entire parameter space is then analysed in Sec.~\ref{ideal}.  In Sec. \ref{SPCM}, we introduce the limitations of real photon detectors based on single photon counting modules, and consider the implications for the cavity operating regime.  An alternative photon detection scheme based on heterodyne detection is presented in Sec. \ref{Het}.  Section~\ref{noise} discusses the susceptibility of the detection process to frequency noise in the system, for different operating conditions, and finally, in Sec.~\ref{quality}, we consider the conversion of SNR to detector quantum efficiency and other limits to detection quality.

\section{\label{model}Cavity QED Model}
The system we are interested in is illustrated in Fig.~\ref{cavity}(a).  It consists of a single two-level atom with an excited state resonance at $\omega_{a}$ coupled to the TEM$_{00}$ mode of an optical cavity with frequency $\omega_{c}$.

The system is driven with a classical (coherent) field at $\omega_{0}$.  Dissipation occurs via spontaneous decay of the atomic excited state, $\gamma$, and cavity field decay, $\kappa$.  The cavity decay comprises transmission through input and output mirrors as well as scattering losses: $\kappa=\kappa_{in}+\kappa_{out}+\kappa_{loss}$ giving a cavity linewidth (FWHM) of $2\kappa$.

The Hamiltonian for this system is that of the driven Jaynes-Cummings model \cite{CohenTannoudji:1998p5976}.  In a reference frame rotating with the driving field,
\begin{equation}
\label{H}
\hat{H}=\hbar\Delta \hat{a}^{\dagger}\hat{a}+\hbar\theta \hat{\sigma}_{+} \hat{\sigma}_{-} +\hbar g (\tilde{r})(\hat{a} \hat{\sigma}_{+}+\hat{a}^{\dagger}\hat{\sigma}_{-})+\hbar\varepsilon(\hat{a}+\hat{a}^{\dagger}).
\end{equation} 
Here $\hat{a}^{\dagger}$ and $\hat{a}$ are the creation and annihilation operators for the cavity mode, and $\hat{\sigma}_{+}=|+\rangle \langle -|$ and $\hat{\sigma}_{-}=|-\rangle \langle +|$ are the atomic pseudospin (raising and lowering) operators for the two-level atom.

The position-dependent atom-field coupling constant is given by $g(\tilde{r})=g_{0}U(\tilde{r})$ where $U(\tilde{r})$ is the normalised magnitude of the electric field.  For a Gaussian standing-wave of waist size $\mathrm{w}_{0}$ and cavity length $L$,  $U(\tilde{r})=\cos(2\pi z/\lambda)\exp(-r^{2}/\mathrm{w}_{0}^{2})$.  The effective mode volume, integrated over the cavity length, is $V=\pi\mathrm{w}_{0}^{2}L/4$.  The single-photon electric field coupling constant for this mode is $g_{0}\equiv\sqrt{\mu^{2}\omega_{c}/(2\hbar\epsilon_{0}V)}$, where $\mu$ is the dipole moment of the atom aligned in the field.  Provided the cavity does not significantly alter the free space atomic decay rate, $\gamma$ (the Purcell effect), the coupling can also be expressed as $g_{0}=\sqrt{\sigma_{0}c\gamma/V}$ for atomic cross section $\sigma_{0}=3\lambda^{2}/(2\pi)$.

The coupling of the cavity field to external modes is $\varepsilon=\sqrt{n_{0}(\kappa^{2}+\Delta^{2})}$, where $n_{0}$ is the empty-cavity photon number.  $\Delta=\omega_{c}-\omega_{0}$ and $\theta=\omega_{a}-\omega_{0}$ are the cavity-probe and atom-probe frequency detunings respectively.

Expectation values for system operators are determined from the steady-state solution to the master equation:
\begin{eqnarray}
\dfrac{d}{dt}\rho(t)=&&-\dfrac{i}{\hbar}[\hat{H},\rho]+
\kappa(2\hat{a}\rho \hat{a}^{\dagger}-\hat{a}^{\dagger}\hat{a}\rho-\rho \hat{a}^{\dagger}\hat{a})\nonumber\\
&& +\dfrac{\gamma}{2}(2\hat{\sigma}_{-}\rho\hat{\sigma}_{+}-\hat{\sigma}_{+}\hat{\sigma}_{-}\rho-\rho\hat{\sigma}_{+}\hat{\sigma}_{-}).
\end{eqnarray}

We model the cavity mode with a truncated Fock state basis, $|0\rangle$,$|1\rangle$,$|2\rangle$\dots$|m\rangle$, that is valid provided $m$ is significantly larger than the mean intra-cavity photon number.  The result is a set of $2m$ linear equations that are solved to find the steady-state density matrix, $\rho$.

The atom is detected inside the cavity via its influence on the cavity field and subsequently the cavity transmission.  The number of photons detected at the output mirror in a measurement interval, $\tau$, is $N=n\kappa_{out}\tau$, where $n=\langle \hat{a}^{\dagger}\hat{a}\rangle$ is the steady-state intra-cavity photon number.  For an empty cavity, $N_{empty}=n_{0}\kappa_{out}\tau$.  The signal for an atom detection event is the difference in these photon numbers, and, assuming the statistics remain Poissonian during an atom transit, the SNR of the measurement is
\begin{equation}\label{SNR}
S = \dfrac{(N_{empty}-N)}{\sqrt{N_{empty}+N}}.
\end{equation}
The assumption of Poissonian noise can break down in extreme regimes of high finesse and large atom cavity coupling where anti-bunching and squeezing can occur \cite{JCarmichael:2007p3325}.  We will, however, concentrate on regimes, particularly with respect to intra-cavity photon number, where these effects are minimal.

\section{Detection Parameter Space}\label{ideal}
By `parameter space' for single atom detection, we refer to variations in cavity-probe, $\Delta$, and atom-probe, $\theta$, detunings, cavity finesse, and probe power.  In this section we present numerical data for the SNR of atom detection with a maximally coupled atom-cavity system ($g(\tilde{r})=g_{0}$).  We find these data are naturally separated into two broad detection regimes: resonant detection at low probe powers, and non-resonant detection at higher powers.  These regions of parameter space are addressed separately in \ref{Resonant Detection} and \ref{Non-Resonant Detection}.

\subsection{Resonant Detection}\label{Resonant Detection}
Initially we consider the resonant condition, a point in parameter space where all three system frequencies are coincident; $\omega_{0}=\omega_{c}=\omega_{a}$.

The system we model is for the $^{87}$Rb D$_{2}$ ($5^{2}$S$_{1/2}\longrightarrow 5^{2}$P$_{3/2}$) transition ($\lambda=780$nm) with $\gamma=2\pi\times 6$MHz \cite{Steck:2008p11}.  The cavity length is $L=100\mu$m, and mode waist $\mathrm{w_{0}}=20\mu$m.  For a cavity with approximately planar mirrors the described geometry does not cover a significant solid angle around the atom.  Consequently, spontaneous emission is to external modes, rather than the single cavity mode, and the Purcell effect can be ignored.  The system has an atom-field coupling of $g_{0}\sim2\pi\times 26$MHz and a cavity decay rate that scales inversely with finesse.  We have chosen to present data for these parameters because they are in the range of realistic experimental design \cite{Ottl:2005p227}, however, the qualitative results that are presented in this work are common to a wide range of design choices.

The data presented here are for an impedance-matched cavity, where $\kappa_{out}=\kappa_{in}=1/2\kappa$, and the empty-cavity transmission is 100\%.  In an experimental set-up, it is the input probe power, rather than transmitted power, that is kept constant during an atom detection event, so data are parameterised in terms of this input photon flux.  In the results presented, we consider $\kappa_{loss}=0$, a reasonable approximation for cavities of moderate finesse.  We also note that one can immediately gain a factor of $\sqrt{2}$ in the SNR, Eq.~(\ref{SNR}), by using an undercoupled cavity with $\kappa_{in}\ll\kappa_{out} \approx \kappa$.

\begin{figure}[t]
	\includegraphics[width=\columnwidth]{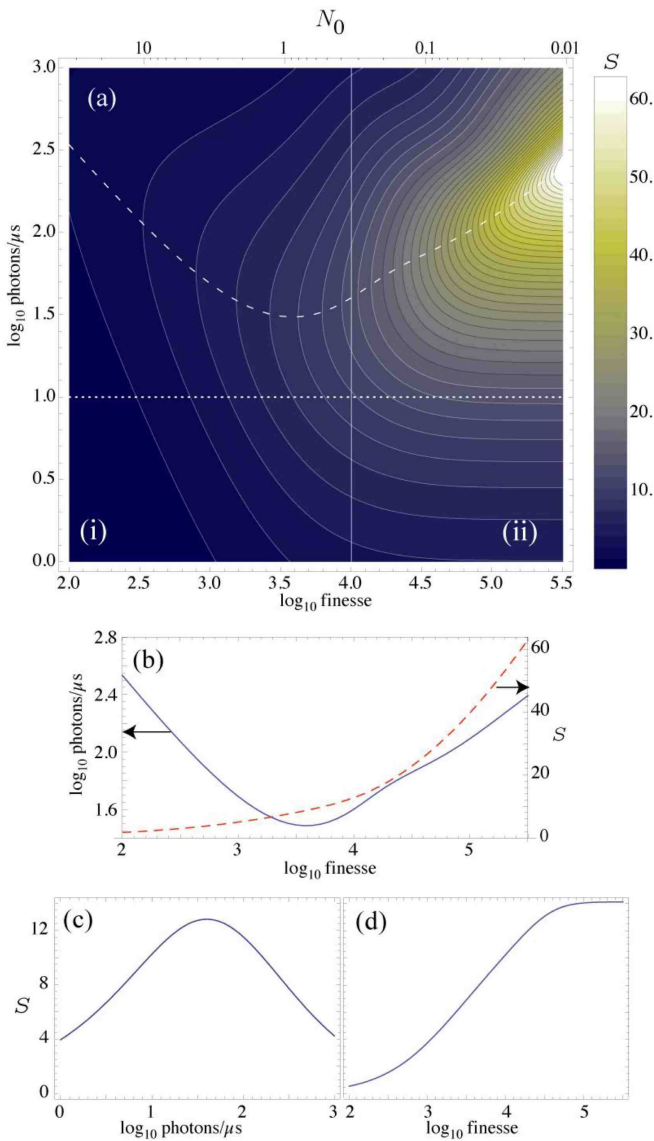}
\caption{\label{FluxvsFinesse} (Colour online) SNR, $S$, for resonant atom detection with ideal photon counting. The cavity has a length of $100\mu$m and waist $20\mu$m.  $\kappa_{in}=\kappa_{out}$, $\kappa_{loss}=0$, $\gamma=2\pi\times6$MHz,  $\tau=20\mu$s.  Figure (b) corresponds to a cross-section along the dashed line in (a), showing the positions of optimum SNR.  The dashed line indicates the SNR and solid line the power as a function of cavity finesse. Figure (c) corresponds to the solid line at $\mathcal{F}=10^{4}$, and (d) to a cross-section along the dotted line in (a) at flux$=10$ photons/$\mu$s.}
\end{figure}

Figure \ref{FluxvsFinesse} shows how the SNR varies with probe power and cavity finesse.  In Fig.~\ref{FluxvsFinesse}(a), the dashed line traces the `position' of maximum SNR.  Figure ~\ref{FluxvsFinesse}(b) shows the cross section along this line, indicating how the optimum SNR improves with finesse and the probe power that is necessary to achieve this optimum.  For a given finesse, there is a clear optimum power at which to operate, shown for $\mathcal{F}=10^{4}$ in Fig.~\ref{FluxvsFinesse}(c).

At $\mathcal{F}\sim~3000$, there is a transition in the system's behaviour; as this value is approached from the low-finesse side, the optimum SNR occurs at lower probe power, while in the limit of high-finesse, the reverse is true, and increasing cavity finesse beyond $3000$ requires increases in power to achieve the maximum SNR.  The transition can be understood in context of the critical atom number, $N_{0}$, and we can separate the plot into regions of (i)~high and (ii)~low critical atom number.

\subsubsection{High Critical Atom Number: Atom as a Saturable Absorber}\label{absorber}
In the low-finesse limit, the critical atom number is large ($>1$), so multiple atoms are necessary to substantially influence the cavity transmission.  An equivalent statement is to say that in this regime the effect of a single atom is only perturbative. In this limit, the atom can be modelled classically as a saturable absorber with absorption cross-section, $\sigma$, that scales, on resonance, as \[\sigma=\dfrac{\sigma_{0}}{I/I_{sat}+1}.\]$I$ is the intensity of light incident on the atom and $I_{sat}$ is the atomic saturation intensity.

We can describe the detection process as follows.  In free space the detected signal, $s$, is a measure of the atom's effect on a photon beam, and is proportional to the photon flux, $F$, and the ratio of the atomic cross-section to the beam area, $A$:
 \[s=F\dfrac{\sigma}{A}.\]

With the atom and probe beam inside a cavity, several changes are observed.  Each photon now passes the atom more than once, effectively increasing $F$.  The atom therefore has a greater chance of absorbing each photon, and has a more significant effect on the probe beam when the absorption happens inside a cavity.  As the finesse increases, so too does the number of round trips for each photon before it decays from the cavity mode, so the signal improves with finesse.

A second effect of the intensity amplification by the cavity, is a reduction in atomic cross-section.  Increases in finesse therefore mean that the external probe power necessary to saturate the atom (inside the cavity) is reduced, and the maximum SNR requires lower power as the finesse increases, in accordance with the dashed line in region (i) of Fig.~\ref{FluxvsFinesse}(a).

\subsubsection{Low Critical Atom Number:  Coupled Resonators}
In contrast with the low-finesse regime, the high-finesse regime requires increasing probe powers with increasing finesse in order to achieve the maximum SNR.

In region (ii) of Fig. \ref{FluxvsFinesse}(a) the critical atom number, $N_{0}$, is less than one, so a single atom significantly influences the system.  The smaller $N_{0}$ becomes, the more significant an effect a single atom will have.  In this regime, we consider the quantum mechanical model in more detail.  That model considers the coupling of two resonators, the bare two-level atom and the empty cavity, giving coupled (dressed) states, $|n+\rangle$ and $|n-\rangle$, that are linear combinations of the uncoupled states: $|n+\rangle=\sin{\alpha}|n-1\rangle|e\rangle+\cos{\alpha}|n\rangle|g\rangle$ and $|n-\rangle=\cos{\alpha}|n-1\rangle|e\rangle-\sin{\alpha}|n\rangle|g\rangle$.  The Jaynes-Cummings energy spectrum of these modes has eigenenergies given by
\begin{subequations}
\begin{equation}\label{eigenenergy}
E(n)_{\pm}=\hbar\omega_{c}(n-1/2)\pm\dfrac{1}{2}\hbar\sqrt{4g^{2}n+(\Delta-\theta)^{2}},
\end{equation}
that reduce, in the case of resonant atom and cavity driving, $\Delta=\theta=0$, to
\begin{equation}
E(n)_{\pm}=\hbar\omega_{0}(n-1/2)\pm\hbar g\sqrt{n}.\label{resonantE}
\end{equation}\end{subequations}
A good review of the dressed states for this resonant condition is given in Ref.~\cite{Alsing:1991p10}.

\begin{figure}[t]
\includegraphics[width=0.65\columnwidth]{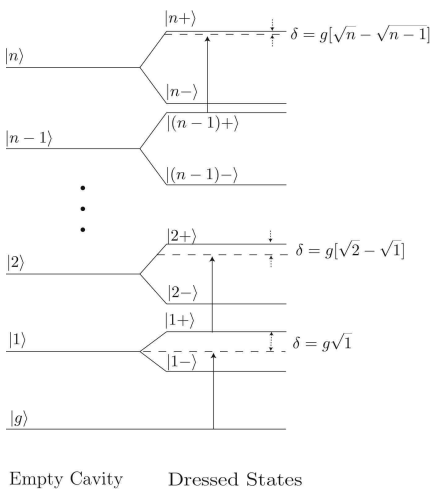}
\caption{\label{eigen} Energy spectrum for resonant dressed states.  Detunings reduce for high levels of excitation.}
\end{figure}

Figure~\ref{eigen} shows the mode-splitting of the dressed atom-cavity system.  When the driving laser is resonant with the uncoupled system, it is detuned from the dressed state resonances in an intensity-dependent manner.  The detuning, $\delta(n)$, is found by considering the addition of a photon from the probe beam ($\hbar\omega_{0}$) to the excited modes in the coupled system:
\begin{eqnarray}
\delta(n)
=&&E(n+1)_{\pm}-[E(n)_{\pm}+\hbar\omega_{0}]\nonumber\\
=&&\pm\hbar g(\sqrt{n+1}-\sqrt{n})\nonumber
\end{eqnarray}
The detuning is largest at low power.  As $n$ increases, the difference between $\sqrt{n+1}$ and $\sqrt{n}$ is reduced ($\sqrt{n+1}-\sqrt{n}\rightarrow 0$ as $n\rightarrow\infty$), so at sufficiently high excitation $\delta(n)$ vanishes and the resonant condition of the uncoupled system is recovered.  Consequently, in the high-power limit, there is no discernible difference between the transmission of an empty cavity and the cavity during an atom transit, and the signal is lost.  It might therefore be expected that the best SNR is obtained at low excitation, however the relative shot noise improves with increasing photon flux, so in fact the optimum SNR occurs at an intermediate power that is a compromise of these limits.

\begin{figure}[t]
	\includegraphics[width=\columnwidth]{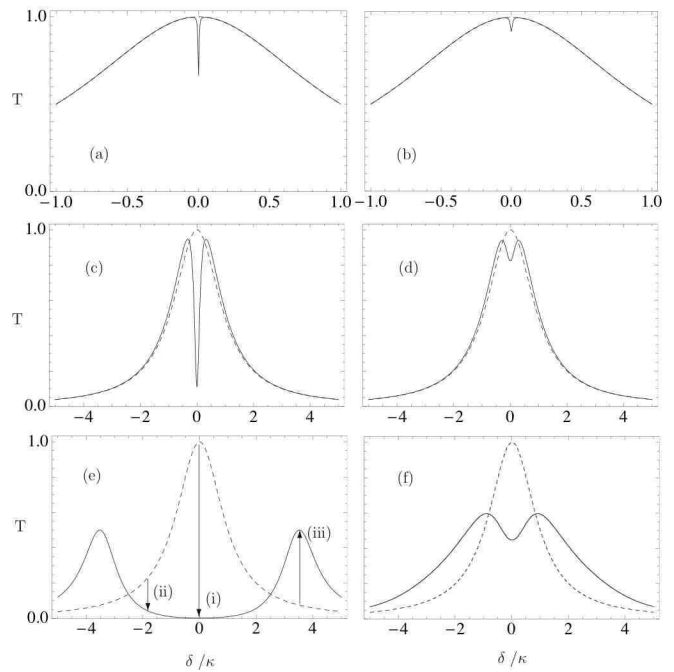}
\caption{\label{mode splitting}Relative photon transmission, T, for the atom-cavity system as a function of probe detuning.  Dashed traces are for the empty cavity and solid lines represent the dressed modes. The cavity finesse $\mathcal{F}$ is (a,b) $10^{3}$, (c,d) $\mathcal{F}=10^{4}$ and (e,f) $10^{5}$. The left plot for each value of $\mathcal{F}$ is for a driving flux of $1$ photon/$\mu$s, the right is for 200 photons/$\mu$s.}
\end{figure}

$\delta(n)$ must be considered in the context of the system decay rates that determine the linewidth of the dressed modes.  Figure~\ref{mode splitting} shows the mode-splitting for $\mathcal{F}=10^{3}$, $10^{4}$ and $10^{5}$, at probe powers of $1$ and $200$ photons/$\mu$s.

In a system with large decay rates, or low finesse (Fig.~\ref{mode splitting}(a)), the dressed state resonant frequencies are not distinctly different from the empty cavity resonance.  Since there is little contrast, the signal is lost easily with increases in intra-cavity photon number (Fig.~\ref{mode splitting}(b)).  As the finesse increases, the dressed state energies become more distinct from the uncoupled resonance (Fig.~\ref{mode splitting}(e)).  In the high-finesse regime, the probe power must be increased considerably before the high-intensity detuning limit ($\sqrt{n+1}-\sqrt{n}\rightarrow 0$) comes into effect.  Consequently, increasing finesse requires increasing probe powers to achieve the maximum SNR, as indicated in region (ii) of Figs.~\ref{FluxvsFinesse}(a) and (b).

If we choose to limit the probe power with which the cavity is driven (the flux limit of single photon counting modules is discussed in Sec.~\ref{SPCM}), then the SNR no longer improves with finesse, as shown in Fig.~\ref{FluxvsFinesse}(d).  In this high-finesse regime, the mode-splitting results in a complete black-out of cavity transmission during an atom transit (shown as the long arrow (i) in Fig.~\ref{mode splitting}(e)).  Further increases in finesse narrow the resonances but cannot improve the signal that is already maximised.  In this regime, the SNR only improves by increasing the probe power.

Although the regimes of high and low critical atom number (regions (i) and (ii) respectively) are best understood in quite different ways, they do have one important feature in common: the best SNR occurs around the point of atomic saturation. In the case of high critical atom number, keeping the circulating power inside the cavity near the saturation intensity means reducing the input power as the finesse increases, as described in section \ref{absorber}.  In the low critical atom number regime, the atom detunes the cavity from the driving field.  Consequently much of the probe light incident on the cavity is reflected and the atom does not experience an enhancement in circulating power due to the cavity.  In order to saturate the atom in this regime, ever increasing amounts of power are required as the cavity finesse is increased.

\subsection{Non-Resonant Detection}
\label{Non-Resonant Detection}
By detuning the probe, we open the possibility of both `positive' and `negative' signals.  The sign is arbitrarily defined by Eq.~(\ref{SNR}), and is not important; both positive and negative signals have been observed in previous cavity detection work \cite{Hood:1998p5016}.  The arrows on Fig.~\ref{mode splitting}(e) show examples of the sign and magnitude of the signal for various detunings.  Arrows pointing down are an indication of positive signals, since they correspond to decreases in cavity transmission, while an upwards pointing arrow implies an increase in transmission during an atom detection event.  For a resonant probe (i), the signal is always seen as a reduction (`dip') in transmitted power, since the atom effectively detunes the system from resonance.  A dip may also be observed if the probe is detuned from the empty cavity (ii).  Alternatively, if the probe is detuned to a position corresponding to a resonance of the dressed system, the atom will bring the system onto resonance with the detuned probe, resulting in an increase (`peak') in the power transmitted (iii).

Combining the choice of probe power with the possibility of detunings dramatically increases the parameter space for atom detection.  We therefore limit the remaining discussion to the cavity design already introduced (cavity length $L=100\mu$m, and waist $\mathrm{w_{0}}=20\mu$m), and consider a modest finesse $\mathcal{F}=10^{4}$.  Additional data for this cavity and a different cavity are available online~\cite{footnote2}.

\begin{figure}[t!]
		\includegraphics[width=1\columnwidth]{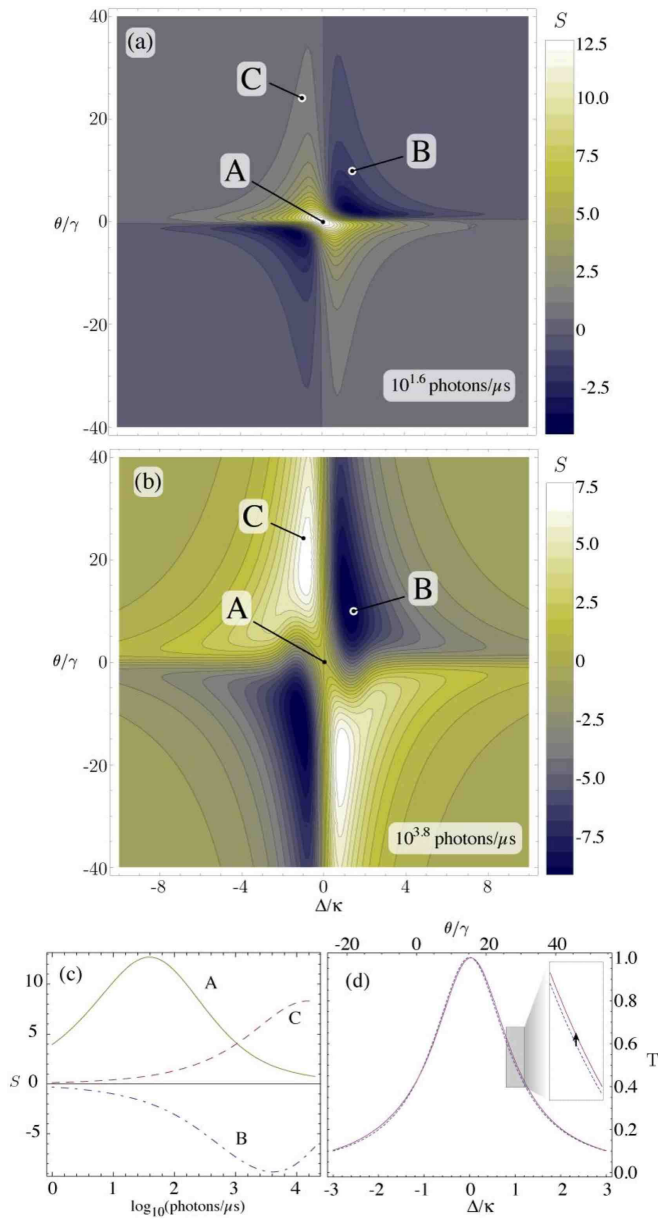}
\caption{\label{detuning} (Colour online) SNR, $S$, for a range of cavity and atom detunings. $\mathcal{F}=10^{4}$, other parameters as in Fig. \ref{FluxvsFinesse}.  The probe power is (a) $\sim40$ photons/$\mu$s, (b) $\sim6300$ photons/$\mu$s.  (c) SNR as a function of probe power positions in detuning space marked A ($\Delta=0\kappa, \theta=0\gamma$), B ($\Delta=\kappa, \theta=10\gamma$) and C ($\Delta=-\kappa, \theta=25\gamma$) in (a) and (b).  (d) Normalised cavity transmission, T, at a probe power of $10^{4}$ photons/$\mu$s.  The atom shifts the cavity resonance from the dashed to solid line, giving a signal represented with the short arrow.} 
\end{figure}

Figure \ref{detuning} presents a selection of data for the SNR, with the resonant condition of Sec.~\ref{Resonant Detection} represented by the position A in the centre of Figs. \ref{detuning}(a) and (b).  Other positions correspond to non-zero detunings.  At low probe power, the best SNR occurs on-resonance and for higher powers, the maximum shifts to detuned operating conditions.  Figure~\ref{detuning}(c) shows the SNR as a function of power for positions A, B and C in Fig.~\ref{detuning}(a) and (b).

The `global maximum' for the SNR (ie: the best SNR obtainable over the whole parameter space) is the peak of $\sim13$ that occurs on resonance, A.  Nonetheless, it may be advantageous to work with red-detuned ($\theta>0$) conditions, due to the benefits of the dipole force that can be used to manipulate the atom's position in the cavity \cite{Ottl:2006p175}.

The mode-splitting shown in Fig.~\ref{mode splitting} is a useful picture for the dressed states of a resonant system.  It is also a good description of the dressed states of a system with equal detunings, $\theta=\Delta$, accessed when the probe laser frequency is scanned but the cavity is kept resonant with the atom ($\omega_{c}=\omega_{a}$), but for more general detunings and at high power, the split modes do not provide useful intuition for the atom's influence on the system.

Figure~\ref{detuning}(d) shows the output flux for the empty cavity and the coupled atom-cavity system at a driving flux of $10^{4}$ photons/$\mu$s with the data normalised to the empty cavity transmission.  The probe frequency is scanned giving atom-probe and cavity-probe detunings indicated on the top and bottom axes.  For these detuning and power conditions, the atom has a dispersive effect, shifting the cavity resonance as shown by the solid line in Fig.~\ref{detuning}(d).  At the resonant peak there is almost no change in transmitted power, since here the amplitude gradient is zero, but on the side of the resonance the small frequency shift means a signal is observed, indicated with the short arrow in the inset of \ref{detuning}(d).  The signal relative to the empty cavity transmission is small, but since it occurs at high probe power the relative shot noise is also small so the SNR can be large.  In \ref{detuning}(c), the magnitude of the optimum SNR at position B is $\sim9$.  This is not as good as the SNR achievable on-resonance but is still very satisfactory.

Work by Horak \textit{et al.} \cite{Horak:2003p31} considered the SNR for far-detuned detection in a different way.  The dispersive interaction of the atom with the cavity field can be measured as a phase shift, rather than a variation in transmission amplitude.  The phase angle measures the difference between the phase of the cavity transmission and the driving laser.  For the empty cavity, the phase angle is zero at the resonant frequency and it is here that the phase gradient is largest.  The frequency shift that the atom induces therefore has the greatest effect on the phase at the transmission peak, rather than side of the resonance where the amplitude gradient is maximum.

\section{Signal Detection Practicalities}
\subsection{\label{SPCM}Single Photon Counting Modules}
In Sec. \ref{ideal} we presented data for an ideal atom detection system where all the transmitted photons at the cavity output mirror are detected.  In practice, this will never be the case, and optimisation of atom detection is critically influenced by the photon measurement process.

A typical single photon counting module is an avalanche photodiode (APD), for example SPCM-AQR-14 \cite{APD}.  This type of device has been employed in experiments by several research groups \cite{Ottl:2005p227, Munstermann:1999p25, Hu:1994p19}.  The quantum efficiency (QE) of an APD is typically around $50$\% at $780$nm, but is non-linear with power.  We will, however, continue to assume an efficiency of 50\%, noting that this generous value is limited to low photon flux.  An acceptable incident photon flux limit is about $20$ photons$/\mu$s, giving an APD count rate of $10$ photons$/\mu$s.  The net result of 50\% efficient detection is a reduction by a factor of $\sqrt{2}$ in the SNR.  The detector flux limit means that, even for the moderate finesse of $\mathcal{F}=10^{4}$ considered here, we cannot reach the probe power required for optimal detection.

\subsection{\label{Het}Heterodyne Detection}
\begin{figure}[b]
\includegraphics[width=\columnwidth]{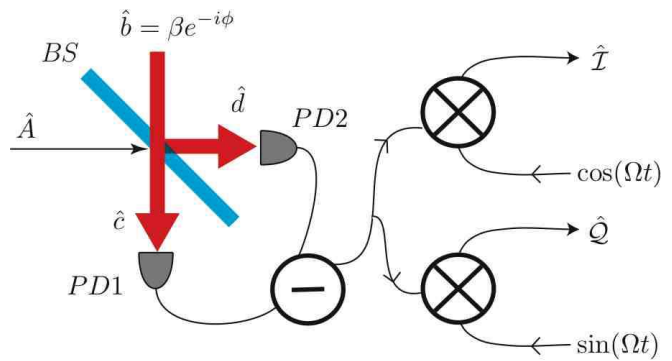}
\caption{\label{het fig} (Colour online) Schematic representation of heterodyne photodetection.}
\end{figure}

So far we have considered the detection of cavity transmission by direct photon counting.  Saturation of real single photon detectors means we are obliged to limit the probe power, and consequently cannot access the optimum SNR of an ideal atom detector.  An alternative is to use heterodyne detection that does not saturate at the probe power discussed here~\cite{Mabuchi:1996p165}.  A possible set-up is indicated in Fig. \ref{het fig}.

Input fields $\hat{A}$ and $\hat{b}$ are combined on a 50-50 beam-splitter.  Here, $\hat{A}=\hat{a}\sqrt{\kappa_{out}\tau}$ is the field at the cavity output mirror and $\hat{b}=\beta e^{-i\phi}$ is that of a strong coherent local oscillator, whose frequency is shifted from the probe laser by $\Omega$.  The phase of the local oscillator can then be expressed as $\phi=\Omega t+\varphi$, where $\Omega$ is the explicit frequency difference, and $\varphi$ is an arbitrary phase difference.  We require the heterodyne signal to detect the influence of the atom on the transmitted probe power.  The process must be independent of any variations in $\varphi$.  To avoid an involved locking procedure, mixing the signal with appropriate $\sin(\Omega t)$ and $\cos(\Omega t)$ components removes the phase dependence as follows:

The beam-splitter output fields, $\hat{c}=1/\sqrt{2}(\hat{A}+\hat{b})$ and $\hat{d}=1/\sqrt{2}(\hat{A}-\hat{b})$, 
are measured as currents at photodiodes PD1 and PD2:
 \[\hat{c}^{\dagger}\hat{c}=\dfrac{1}{2}(\hat{A}^{\dagger}\hat{A}+\hat{b}^{\dagger}\hat{b}+\hat{A}^{\dagger}\hat{b}+\hat{b}^{\dagger}\hat{A}),\]
\[\hat{d}^{\dagger}\hat{d}=\dfrac{1}{2}(\hat{A}^{\dagger}\hat{A}+\hat{b}^{\dagger}\hat{b}-\hat{A}^{\dagger}\hat{b}-\hat{b}^{\dagger}\hat{A}).\]
Subtracting these photocurrents we find
\[\hat{c}^{\dagger}\hat{c}-\hat{d}^{\dagger}\hat{d}=\beta(\hat{A}^{\dagger}e^{-i\phi}+\hat{A}e^{i\phi}).\]
Mixing the signal with sine and cosine functions that oscillate at $\Omega t$ gives terms 
\begin{eqnarray}
\hat{\mathcal{I}}
=&&\cos(\Omega t)\times\beta(\hat{A}^{\dagger}e^{-i(\Omega t+\varphi)}+\hat{A}e^{i(\Omega t+\varphi)})\nonumber\\
=&&(\beta /2)(\hat{X}^{-\varphi}+\hat{X}^{-(2\Omega t +\varphi)}),
\label{cos}
\end{eqnarray}
\begin{eqnarray}
\hat{\mathcal{Q}}
=&&\sin(\Omega t)\times\beta(\hat{A}^{\dagger}e^{-i(\Omega t+\varphi)}+\hat{A}e^{i(\Omega t+\varphi)})\nonumber\\
=&&(\beta /2)(\hat{X}^{-(\varphi +\pi /2)}+\hat{X}^{-(2\Omega t +\varphi -\pi /2)})\label{sin}
\end{eqnarray}
where we have used the amplitude, $\hat{X}^{+}=(\hat{A}^{\dagger}+\hat{A})$, and phase, $\hat{X}^{-}=-i(\hat{A}-\hat{A}^{\dagger})$, quadratures of $\hat{A}$ and expressed the results with $\hat{X}^{\vartheta}=\hat{X}^{+}\cos(\vartheta)+\hat{X}^{-}\sin(\vartheta)$.  Eqs. (\ref{cos}) and (\ref{sin}) are used to generate the final measurement,
\begin{eqnarray}\label{measurementeq}\langle \hat{\mathcal{I}}\rangle^{2}+\langle\hat{\mathcal{Q}}\rangle^{2}
=&&(\beta^{2}/4)[\langle\hat{X}^{+}\rangle^{2}+\langle\hat{X}^{-}\rangle^{2}]\nonumber\\
=&&\beta^{2}\langle\hat{A}\rangle\langle\hat{A}^{\dagger}\rangle.
\end{eqnarray}

Terms with $2\Omega t$ dependence in $\hat{\mathcal{I}}$ and $\hat{\mathcal{Q}}$ are vacuum terms  \cite{Yuen:1980p1160,Yuen:1983p1290}.  They do not contribute to the signal but add to the noise which is determined by examining the variance of $\hat{\mathcal{I}}$ and $\hat{\mathcal{Q}}$.  
The variance $V_{\hat{\mathcal{I}}}\equiv(\Delta\hat{\mathcal{I}})^{2}$ is given by the variances of  the measured and vacuum fields: $V^{\varphi}=\langle (X^{-\varphi})^{2}\rangle-\langle X^{-\varphi}\rangle^{2}$, and $V^{2\Omega t+\varphi}=\langle (X^{-(2\Omega t+\varphi)})^{2}\rangle-\langle X^{-(2\Omega t+\varphi)}\rangle^{2}$ respectively.
\begin{eqnarray}
V_{\hat{\mathcal{I}}}
=&&\langle\hat{\mathcal{I}}^{2}\rangle-\langle\hat{\mathcal{I}}\rangle^{2}\nonumber\\
=&&(\beta^{2}/4)[V^{\varphi}+V^{2\Omega t+\varphi}]\nonumber\\
=&&\beta^{2}/2,\nonumber
\end{eqnarray}
with an identical result for $V_{\hat{\mathcal{Q}}}$.  Here we have assumed the field remains coherent, so the variances of the measured and vacuum fields are both one: $V^{\varphi}=V^{2\Omega t+\varphi}=1$.

The total noise on the measurement is
\begin{eqnarray}\label{noiseeq}\Delta(\langle \hat{\mathcal{I}}\rangle^{2}+\langle\hat{\mathcal{Q}}\rangle^{2})
=&&\sqrt{(2\langle\hat{\mathcal{Q}}\rangle\Delta\hat{\mathcal{Q}})^{2}+(2\langle\hat{\mathcal{I}}\rangle\Delta\hat{\mathcal{I}})^{2}}\nonumber\\
=&&\beta^{2}\sqrt{2\langle\hat{A}\rangle\langle\hat{A}^{\dagger}\rangle}.
\end{eqnarray}

The expressions for the measurement (Eq.~(\ref{measurementeq})) and noise (Eq.~(\ref{noiseeq})) replace $N=\kappa_{out}\tau\langle\hat{a}^{\dagger}\hat{a}\rangle$ in Eq.~(\ref{SNR}), and a similar approach is taken for the empty cavity where $\langle\hat{a}\rangle_{0}\langle\hat{a}^{\dagger}\rangle_{0}=\langle\hat{a}^{\dagger}\hat{a}\rangle_{0}=n_{0}$.  The SNR for atom detection with a heterodyne set-up is therefore of a slightly different form to that of direct detection;
\begin{eqnarray}
S_{het}=\dfrac{\sqrt{\kappa_{out}\tau}(n_{0}-\langle\hat{a}\rangle\langle\hat{a}^{\dagger}\rangle)}{\sqrt{2(n_{0}+\langle\hat{a}\rangle\langle\hat{a}^{\dagger}\rangle)}}.
\end{eqnarray}
The noise includes the usual factor of $\sqrt{2}$ of heterodyne measurements \cite{Yuen:1980p1160,Yuen:1983p1290}.  However, the SNR for heterodyne detection is not necessarily smaller by $\sqrt{2}$ than for direct detection, since the signal and noise now both contain expectation values of different quantum operators.

\begin{figure}[t]
	\includegraphics[width=\columnwidth]{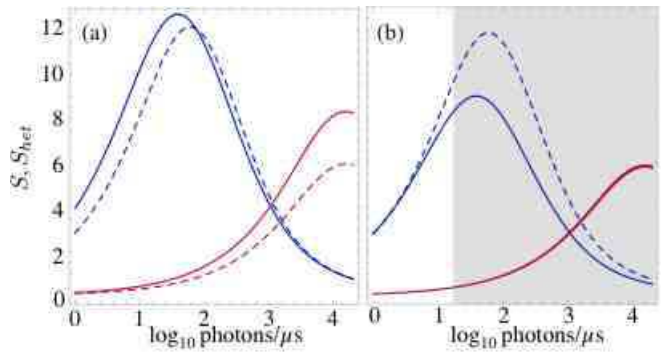}
\caption{\label{comparison} (Colour online) A comparison of the SNR, $S$ (solid) and $S_{het}$ (dashed), for atom detection with (a) ideal photon counting and ideal heterodyne detection and (b) an APD with $50$\% QE and a saturation flux of $20$ photons$/\mu$s, and a realistic heterodyne set up using $95\%$ efficient photodiodes.  Traces at high power are for detection at the detuned position, C, in Fig.~\ref{detuning}. Parameters as for Fig. \ref{detuning}.}
\end{figure}

Figure~\ref{comparison} shows a comparison of the detection schemes.  The solid curves are for detection of $\langle\hat{a}^{\dagger}\hat{a}\rangle$, using an APD, and the dashed curves represent heterodyne detection; $\langle\hat{a}^{\dagger}\rangle\langle\hat{a}\rangle$.  Traces that peak at low probe power are for the resonant condition and those that peak at higher power represent detection in the detuned region marked C in Fig.~\ref{detuning}.  In Fig. \ref{comparison}(a), the comparison is between ideal direct detection and ideal heterodyne detection, using photon detectors with $100$\% QE.  For resonant detection the maximum SNR is similar in both detection schemes, although the probe power that is necessary to achieve it varies somewhat.

Figure~\ref{comparison}(b) includes the quantum efficiencies of real detectors; APD efficiency is $50$\% and the photodiodes used in the heterodyne set-up are assumed to be $\sim 95$\% efficient.  The shaded region indicates the probe powers that are inaccessible to real APDs due to their flux-limit.  These realities show a significant difference between the two detection schemes.  On-resonance, the best SNR for the flux-limited APD is $\sim8$, whereas using heterodyne detection at a higher power can achieve a SNR of $\sim12$.

For non-resonant detection the maximum SNR of APD and heterodyne detection are in principle very similar.  However, since the maximum occurs at high power, APDs are less appropriate for detection in that regime.

\subsection{Noise Susceptibility}\label{noise}
The SNRs presented in our data are determined using photon statistics that are always shot-noise limited.  The ability to achieve this depends on the stability of the system as well as the choice of operating regime.  The operating condition least sensitive to frequency noise is the resonant position ($\Delta=0$) at the peak of the cavity transmission line since here small changes in detuning have little effect on the transmitted power.  Detuned detection at the side of the transmission line, as suggested in Fig.~\ref{detuning}(d), is more sensitive to frequency noise since here the gradient on the amplitude is large so even small fluctuations in cavity detuning can significantly influence the power transmitted.

In the previous sections, two broad detection regimes have been presented.  Section~\ref{Resonant Detection} covered the SNR behaviour for resonant detection where the observed signal is seen as a significant drop in amplitude at the transmission peak due to Jaynes-Cummings mode splitting.  In Sec.~\ref{Non-Resonant Detection} the signal is observed from a point on the side of the transmission line where the signal is due to a small frequency shift of the line that causes a change in transmitted light at the detection frequency.

Frequency noise enters the system via variations in the probe laser frequency as well as the cavity length, so, to eliminate noise, these components must be extremely stable.  The cavity design emphasised in this work has a length of $100\mu$m and finesse $\mathcal{F}=10^{4}$, giving a linewidth of almost $1$GHz.  Assuming frequency locking on the order of $1$MHz, amplitude fluctuations at the position of the signal marked by the short arrow in Fig.~\ref{detuning}(d) are roughly one part in $1000$.  The transmitted power at this position is $\sim5000$ photons/$\mu$s, with shot noise of about one part in 100, so the amplitude fluctuations due to frequency-locking limitations are well below the shot noise limit.  For longer cavities or for cavities with higher finesse, however, the transmission linewidth decreases, so the relative frequency stability drops significantly.

For some operating conditions and cavity designs the influence of frequency noise may therefore become difficult to eliminate.  In such cases, if the noise can be measured, it can be simply subtracted from the signal.  Measuring the noise via an error signal is an ideal solution, since a possible mechanism for locking the laser, atom and cavity set-up involves the use of a far-detuned stabilisation laser and `transfer cavity' in addition to the probe beam.  This technique is described in Ref.~\cite{Ottl:2006p175}.  The second laser could provide the necessary error signal to be subtracted from the measurement.

\subsection{Detection Efficiency}\label{quality}
\subsubsection{Discriminator Position}
\label{Discriminator}
Having identified regions of parameter space that maximise the SNR, a further question for signal analysis regards the separation of a detection event from the shot-noise of the empty cavity transmission, indicated with the dotted `discriminator' line in Fig.\ref{cavity}(b).

Variations in the number of photons counted in a measurement interval of $20\mu$s, for an empty cavity, are photon shot-noise fluctuations.  So far, we have considered the same to be true of the number of counts from the cavity during an atom transit, and the two Poissonian distributions are related by the SNR.  The QE and false count rate of the atom detection depend on the value chosen to distinguish between these photon distributions.  For a resonant signal, where the detected photon number during an atom transit is less than that of an empty cavity, raising the discriminator increases the QE since it includes more of the distribution of `signal' counts, but more of the empty-cavity counts are also included, so the false counts increase.

\begin{figure}[!ht]
\includegraphics[width=\columnwidth]{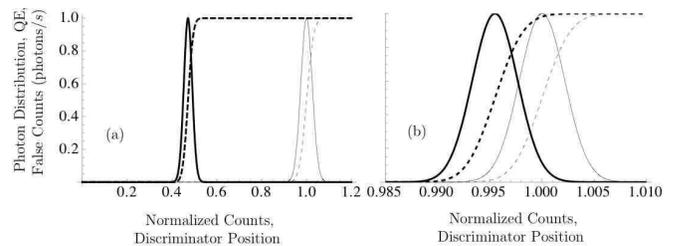}
\caption{\label{QE}Variation in QE and false detection count rate with discriminator position.  Solid traces are the normalised photon counts for the empty cavity (fine) and the cavity containing an atom (thick).  Dashed traces correspond to the QE (thick) and false counts (fine) as the position of the discriminator is varied.  The SNR is (a)$\sim17$, (b)$\sim1.5$.}
\end{figure}
The QE (false count rate) is the integration of the Poissonian photon distribution about $N$ ($N_{empty}$) from zero up to the value at the discriminator position.  Figure~\ref{QE} shows how the QE and false counts vary as the discriminating value is changed, for high and low SNR (Figs.~\ref{QE}(a) and (b) respectively).

When the empty-cavity and signal count rates are well separated, as in Fig.~\ref{QE}(a), it is clear the discriminator must lie in the range between the two photon distributions.  Provided this range is large enough, a position can be assigned (for example) at least three standard deviations from each distribution (3$\sigma_{N_{empty}}$ below $N_{empty}$ and 3$\sigma_{N}$ above $N$), giving a QE greater than $99$\%, and probability of false counts less than $1$\%.  As the number of standard deviations is increased, the detection quality improves.  The necessary SNR depends on the desired detection quality; for a discrimination position $p$ standard deviations ($p\sigma$) from the centre of each photon count distribution, the minimum SNR required is $\sqrt{2}p$.

In Fig.~\ref{QE}(b), for low SNR, the photon count distributions are not well separated and it is difficult to find a position for the discriminator that is a good compromise between QE and false counts.  Such statistics do not represent ideal detection conditions.  However, the false counts must be considered in context of the atom flux.  For a large (true) atom count rate, a high false count rate becomes more acceptable.  In such a case, the discriminator may be positioned to include a significant fraction of the empty-cavity count distribution about $N_{empty}$.

\subsubsection{Limits to Efficiency}
It is important to note the limitations of our model with respect to detection efficiency.  We have considered maximal coupling between the atom and cavity field ($g(\tilde{r})=g_{0}$), neglecting any variation in coupling strength (and corresponding signal strength) that occurs when the atom transits the cavity away from the intensity maximum of the light field.  In a real system, this variation critically effects the QE.  If the low field intensity around the nodes of a standing-wave cavity can be avoided, the coupling strength seen by each atom can be made substantially more uniform.  This can be achieved in two ways:

\textit{(i)} For a non-resonant red-detuned probe, the electric dipole force can be used to pull the atoms through the intensity maxima of the cavity mode.  The dipole potential is proportional to the field gradient, so the rapid intensity changes of a standing-wave can create a strong force along the cavity axis.  On the basis of our modelling, we expect the regime around point B in Fig.~\ref{detuning} would provide both a strong axial dipole potential and a reasonable SNR that is reduced by a factor of about 1.4 compared to best case resonant detection.  There is also a weaker dipole force in the radial direction due to the Gaussian beam profile.  In principle, this could increase the effective width of the atom detector.  Assuming reasonable atom speeds of about 1~$m/s$, however, the radial dipole force is not strong enough to substantially influence the atom trajectories.  This is true both for the regimes considered here and in other work \cite{Ottl:2006p175}.

\textit{(ii)} Only linear cavities have axial mode structure.  A travelling-wave ring cavity has a field that is uniform along the cavity axis.  Consequently all atoms transiting the field on-axis generate signals of the same strength.  In this configuration, there is little to be gained from using red-detuned light since the dipole force acting on the atoms will only be in the radial direction, and, as discussed above, unable to significantly influence atom trajectories.  A travelling-wave cavity therefore seems highly suited to resonant detection where the SNR is maximised.  The downside is that the lack of standing-wave structure yields a mode volume twice that of a linear cavity with the same round-trip length.  This reduces $g_0$ by a factor of $\sqrt{2}$. Our modelling indicates that for both resonant and detuned detection with optimum probe power, travelling-wave cavities loose a factor of 1.3 to 1.5 in SNR compared to standing-wave cavities with otherwise identical properties.  It is also problematic that real multi-level atoms need circularly polarised light to drive a closed two-level transition. Due to the birefringence of the dielectric mirrors used in high finesse cavities, planar ring cavities have vertical and horizontal polarisation modes that are non-degenerate. To force a resonant circular polarisation mode would mean some polarisation compensation inserted into the cavity, or a complex 3-D geometry that is symmetric with respect to the linear polarisation modes.

In summary, non-resonant detection allows one to use a linear cavity with a standing wave and higher $g_0$, but one is forced to consider red-detuned detection, which has lower potential SNR than the resonant system.  Alternatively, a ring cavity with no standing wave, is better suited to resonant detection, but comes at a cost of potential SNR due to the reduction in $g_0$, and is in practice difficult to set up.  The end result is that both options give similar performance.

Although there are many experimental details that we have not considered in the present work, our model is still useful for comparing real set-ups.  For example, \"{O}ttl \textit{et al.} detected single atoms in a $^{87}$Rb atom-laser beam using a high finesse ($\mathcal{F}=3.5\times10^{5}$) optical standing-wave cavity, with $L=178\mu$m, $\mathrm{w}_{0}=25.5\mu$m  \cite{Ottl:2006p175}.  Their detection made use of the dipole force to channel the atoms through antinodes of the cavity field, and optimum detection efficiency occurred for detunings of $\theta=3\gamma$ and $\Delta=0.5\kappa$ with a driving photon flux of $70$ photons/$\mu$s.  In our simulations, these cavity parameters and operating conditions suggest an ideal SNR for single atom detection of about 10.  Their cavity is appropriate for many experiments besides single atom counting since it accomplishes strong coupling conditions \cite{Ottl:2005p227}.  The cavity design that we have discussed in this paper has a finesse that is an order of magnitude lower ($\mathcal{F}=10^{4}$) than the cavity presented in the work by \"{O}ttl \textit{et al}.  Nonetheless, we have demonstrated that for our moderate finesse, there exist operating regimes, in both resonant and detuned conditions, where the achievable SNR is as good as that of the higher finesse cavity.  The critical difference between the optimum operating regimes used in Ref.~\cite{Ottl:2006p175} and those shown in this work, is the use of high probe powers necessitating detection with a heterodyne set-up.  

\section{Conclusions}
In this work, we have presented a thorough analysis of single atom detection using optical cavities.  The parameter space considered includes cavity-probe and atom-probe detunings as well as variable probe power, and we have shown that the SNR for single atom detection is critically dependent on the choice of operating regime within this space.

Our modelled data suggest the parameter space be divided into two regimes: resonant and non-resonant detection.  Resonant detection with moderate to high-finesse cavities (systems with low critical atom number) is best described with the Jaynes-Cummings mode-splitting.  Non-resonant detection results in a frequency-pulling of the cavity transmission line.

The best SNR occurs on-resonance.  However, very reasonable SNRs are also available with non-resonant conditions, provided the atom and cavity detunings are chosen wisely and combined with appropriately high probe powers.  With a standing-wave configuration, red-detuned detection brings the benefit of the dipole potential that improves the effective atom-cavity coupling, however, equivalent SNRs are achieved with resonant detection in a travelling-wave ring cavity.

We have shown maximizing the SNR for both resonant and non-resonant conditions requires photon fluxes that are in excess of APD saturation limits, so heterodyne detection is always a more desirable detection technique.  Working in high power regimes means that for a cavity of moderate finesse, $\mathcal{F}=10^{4}$, we can achieve a SNR comparable or better than those achieved in previous experiments using cavities with significantly higher finesse.

\section{Acknowledgements}
We wish to thank Joseph Hope, Andre Carvalho and Ben Weise for their valuable discussions and contribution to the initial coded model.

This work was supported by the Australian Research Council Centre of Excellence for Quantum-Atom Optics and the Defence Science and Technology Organisation.


\begin{thebibliography}{39}
\bibitem{footnote}{For groups interested in accessing the Mathematica code used in our model, correspondence is directed to the indicated author.}
\bibitem{footnote3}{There is some variation in relevant literature for expressions involving the atomic dissipation rate.  The rate, $\gamma$, that we use here, is the rate of decay for the population of the atomic excited state.  It is also common to see this dissipation parameter expressed as the spontaneous emission rate, which is smaller by a factor of two (for example, Ref. \cite{Kimble:1998p6}).}
\end{thebibliography}
\end{document}